\documentclass[%
 reprint,
superscriptaddress,
 amsmath,amssymb,
 aps,
]{revtex4-2}

\usepackage{graphicx}%

\newcommand\cl{\columnwidth}

\begin{document}

% Include your paper's title here

\title{Optimum control strategies for maximum thrust production in underwater undulatory swimming}

% Place the author information here.  Please hand-code the contact
% information and notecalls; do *not* use \footnote commands.  Let the
% author contact information appear immediately below the author names
% as shown.  We would also prefer that you don't change the type-size
% settings shown here.

\author{Li Fu}
 \affiliation{Laboratoire de Tribologie et Dynamique des Systemes, \'Ecole Centrale
de Lyon, CNRS}%Lines break automatically or can be forced with \\
\author{Sardor Israilov}
 \affiliation{Universit\'e C\^ote d'Azur, CNRS, INPHYNI, 17 rue Julien Laupr\^etre, 06200 Nice, France}
\author{Jes\'us S\'anchez-Rodr\'iguez}%
\affiliation{Departamento de F\'isica Fundamental, Universidad Nacional de Educaci\'on a Distancia,  Madrid, 28040, Spain}

\author{Christophe Brouzet}
 \affiliation{Universit\'e C\^ote d'Azur, CNRS, INPHYNI, 17 rue Julien Laupr\^etre, 06200 Nice, France}

 \email{Second.Author@institution.edu}

\author{Guillaume Allibert}
 \affiliation{Universit\'e C\^ote d'Azur, CNRS, I3S, Sophia Antipolis, France}

\author{Christophe Raufaste}
 \affiliation{Universit\'e C\^ote d'Azur, CNRS, INPHYNI, 17 rue Julien Laupr\^etre, 06200 Nice, France}
 \affiliation{IUF, Paris, France}
\author{M\'ed\'eric  Argentina}
 \affiliation{Universit\'e C\^ote d'Azur, CNRS, INPHYNI, 17 rue Julien Laupr\^etre, 06200 Nice, France}
 
\date{\today}% It is always \today, today,

\begin{abstract}
%Fish, cetaceans and many other aquatic vertebrates undulate their bodies to propel themselves through water. Numerous studies on natural, artificial or analogous swimmers are dedicated to revealing the links between the kinematics of body oscillation and the production of thrust for swimming. One open and difficult question concerns the best kinematics to maximize this force for given constraints and how a system  tunes its internal parameters to reach this maximum. To address this challenge, we exploit a biomimetic robotic swimmer to determine the control signal that produces the highest thrust. Using machine learning techniques and intuitive models, we find that this optimal control consists of a square wave function, whose frequency is fixed by the interplay between the internal dynamics of the swimmer and the fluid-structure interaction with the surrounding fluid. We then propose a simple implementation for autonomous robotic swimmers that requires no prior knowledge of systems or equations. This application to aquatic locomotion is validated by 2D numerical simulations.

Fishes, cetaceans, and many other aquatic vertebrates undulate their bodies to propel themselves through water. Swimming requires an intricate interplay between sensing the environment, making decisions, controlling internal dynamics, and moving the body in interaction with the external medium. Within this sequence of actions initiating locomotion, biological and physical laws manifest complex and nonlinear effects,  which does not prevent natural swimmers to demonstrate efficient movement. This raises two complementary questions: how to model this intricacy and how to abstract it for practical swimming. In the context of robotics, the second question is of paramount importance to build efficient artificial swimmers driven by digital signals and mechanics.
In this study, we tackle these two questions by leveraging a biomimetic robotic swimmer as a platform for investigating optimal control strategies for thrust generation. Through a combination of machine learning techniques and intuitive models, we identify a control signal that maximizes thrust production. Optimum tail-beat frequency and amplitude result from the subtle interplay between the swimmer's internal dynamics and its interaction with the surrounding fluid. We then propose a practical implementation for autonomous robotic swimmers that requires no prior knowledge of systems or equations. Direct fluid-structure simulations confirms the effectiveness and reliability of the proposed approach. Hence, our findings bridge fluid dynamics, robotics, and biology, providing valuable insights into the physics of aquatic locomotion

\end{abstract}

                              %display desired
\maketitle

\section{Introduction}

The diversity of shapes and physiologies among multicellular organisms is tremendous, and it can be rationalized by the principles of Darwinian evolution and the various functions required by animals \cite{healy2019animal}. Locomotion is a vital activity for metazoans, as it enables them to fulfill essential functions necessary for survival, such as accessing favorable environments, engaging in reproduction, hunting, and evading predators.

From tadpoles of a few centimeters to whales of 20 meters in length, swimming consists in pushing the water by undulating the body~\cite{childress1981mechanics} which produces a thrust exploiting the inertia of the displaced fluid \cite{gazzola2014scaling}. The kinematics of underwater undulatory swimming appear to be particularly robust in vertebrates, highlighting general physical principles. The wavelength of the body deformation is of the order of the length of the swimmer \cite{videler_fish_1993,DiSanto2021}, while there is on average a factor 0.2 between tail beat amplitude and swimmer length~\cite{bainbridge1958speed, rohr2004strouhal,hunter1971swimming, saadat2017rules, sanchez2023scaling}. The tail beat frequency~$f$ is not fixed for an individual but tunes its swimming speed: the higher the frequency, the higher the speed~\cite{bainbridge1958speed, triantafyllou1993optimal,gazzola2014scaling, saadat2017rules,sanchez2023scaling}. There is evidence that each swimmer can vary its frequency within a frequency band whose range is set by the interplay between the muscle properties and the interaction of the swimmer with its surrounding fluid~\cite{sanchez2023scaling}. Muscles have their own limits in terms of speed of contraction and tension, as represented by Hill's muscle model~\cite{hill1938heat}. Constraints can also be imposed by decision processes that are either spontaneous, through proprioceptive reflexes \cite{pearson1995proprioceptive, williams2013function, sanchez-rodriguez_proprioceptive_2021}, or conscious, for instance through the choice of the activity level~\cite{brett1964respiratory, brett_metabolic_1972}. These decisions drive the gait~\cite{goerig2021convergence} and we can expect different control strategies for an individual swimming at burst speed~\cite{hirt2017general} and the same swimmer exhibiting a swim-and-coast gait in a sustained level of activity~\cite{li_burst-and-coast_2021}.

These considerations are echoed in biomimetic robotic swimmers~\cite{lebastard_reactive_2016, zhu_tuna_2019,sanchez-rodriguez_proprioceptive_2021,thandiackal_emergence_2021,lee_autonomously_2022}. The biological nature of the internal dynamics is here replaced by robotic elements from electronics, mechanics and/or computer science. Developing efficient and fully autonomous artificial swimmers requires finding the appropriate control strategies tailored to specific constraints to achieve optimal performance for a given action. 
Recent years have witnessed the development of various approaches to control soft robot fish, both in simulations and experiments. 
Classical  control techniques such as PI \cite{Zhang2015}, PID \cite{Yu2004}, and robust controllers \cite{Zhang2016} have been employed to improve trajectory tracking performance. 
The emergence of artificial intelligence algorithms has paved the way for novel control approaches designed specifically for soft swimmers. 
Models and simulations are now utilized to explore various machine learning algorithms, achieving high swimming speeds while maintaining trajectory accuracy \cite{Novati2019, Kanso2021, Rajendran2022, Yousseff2022}. 
Regardless of the methodology employed, the quest to achieve the highest swimming speed through robotic fish remains an enduring challenge in the domain of aquatic locomotion. While certain investigations have explored the correlation between swimming speed and tail beat frequency \cite{Epps2009, Zhu2019, Yousseff2022}, none of these prior studies has offered a conclusive comprehension of the optimal control strategy required to attain this objective.

In this article, we present our findings on the most effective scheme for generating maximum thrust in swimming using a single-parameter control system. Initially, we showcase the experimental outcomes achieved by employing deep reinforcement learning (RL) techniques \cite{Sutton2012} to drive a swimming robot. Through these experiments, we identify the optimal control strategy, which involves utilizing a square wave function that alternates between the two extreme values allowed by the controller. 
To gain a deeper understanding of the underlying physics, we develop a theoretical model that further supports and explains the observed results. Subsequently, we rigorously validate our optimized thrust strategy through a comprehensive 2D numerical simulation. By aligning experiments, theory, and numerical simulations, we successfully determine the most efficient approach to achieve maximum swimming speed. This integrated approach provides valuable insights into the best methods for enhancing propulsion in aquatic systems.

\section{Results}\label{secResults}
\subsection{Experiments and machine learning}
\begin{figure}[ht]
    \centering
    \includegraphics[width=\cl]{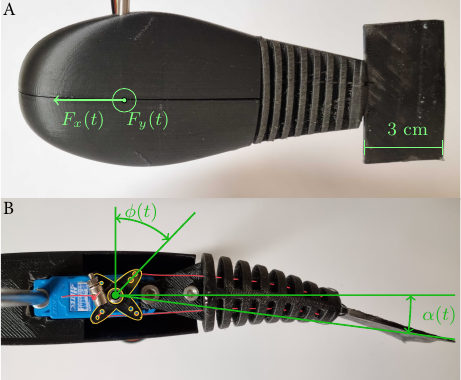}	
    \caption{\label{fig:setup}
            (A) Side view of the robotic fish. (B) Top view without the top fairing. The head is attached to a dual-force sensor. The tail and the fin are printed with a flexible polymer. The body deformation is triggered by the actuation of a waterproof servomotor (blue):  the rotation of its wheel pulls cables (drawn in red) to deform the elastic tail~\cite{gibouin2018study}. The contours of the servomotor wheel are drawn in yellow.
            } 
\end{figure}

We utilized the experimental platform described in \cite{gibouin2018study,sanchez-rodriguez_proprioceptive_2021}, which was equipped with the robotic fish illustrated in Fig.~\ref{fig:setup}. The robotic fish consisted of a deformable skeleton with a fin attached to its end, fabricated through 3D printing using a flexible polymer. To achieve controlled deformation, a servomotor was employed, connected to the skeleton's end via two cables. By rotating the wheel of the servomotor by an angle $\phi(t)$, we induced a deformation that controlled the fin angle $\alpha(t)$. This design closely emulates the functioning of antagonistic muscles found in natural swimmers, responsible for initiating body deformations. 
The robot was immersed inside a water tank and its head was fixed to a force sensor that measured the longitudinal force $F_x(t)$. This force served as a measure of thrust and was, on average, positive when the fish was in the propulsion phase. 

We employed an efficient RL algorithm to identify the optimal control signal that maximized the average thrust. In accordance with the specified details in the Materials and Methods section, we limited the instruction angle $\phi_c(t)$ of the servomotor to vary within the interval $[-\Phi,\Phi]$. The RL algorithm was designed to maximize a reward proportional to the thrust~$F_x(t)$.
The algorithm required inputs to characterize the state of the system at every time step.
Two independent sets of inputs were considered consecutively to probe the robustness of the learning process and ensure convergence toward the optimal control signal. 

\begin{figure}[t]
        \centering
        \includegraphics[clip=true,width=\cl]{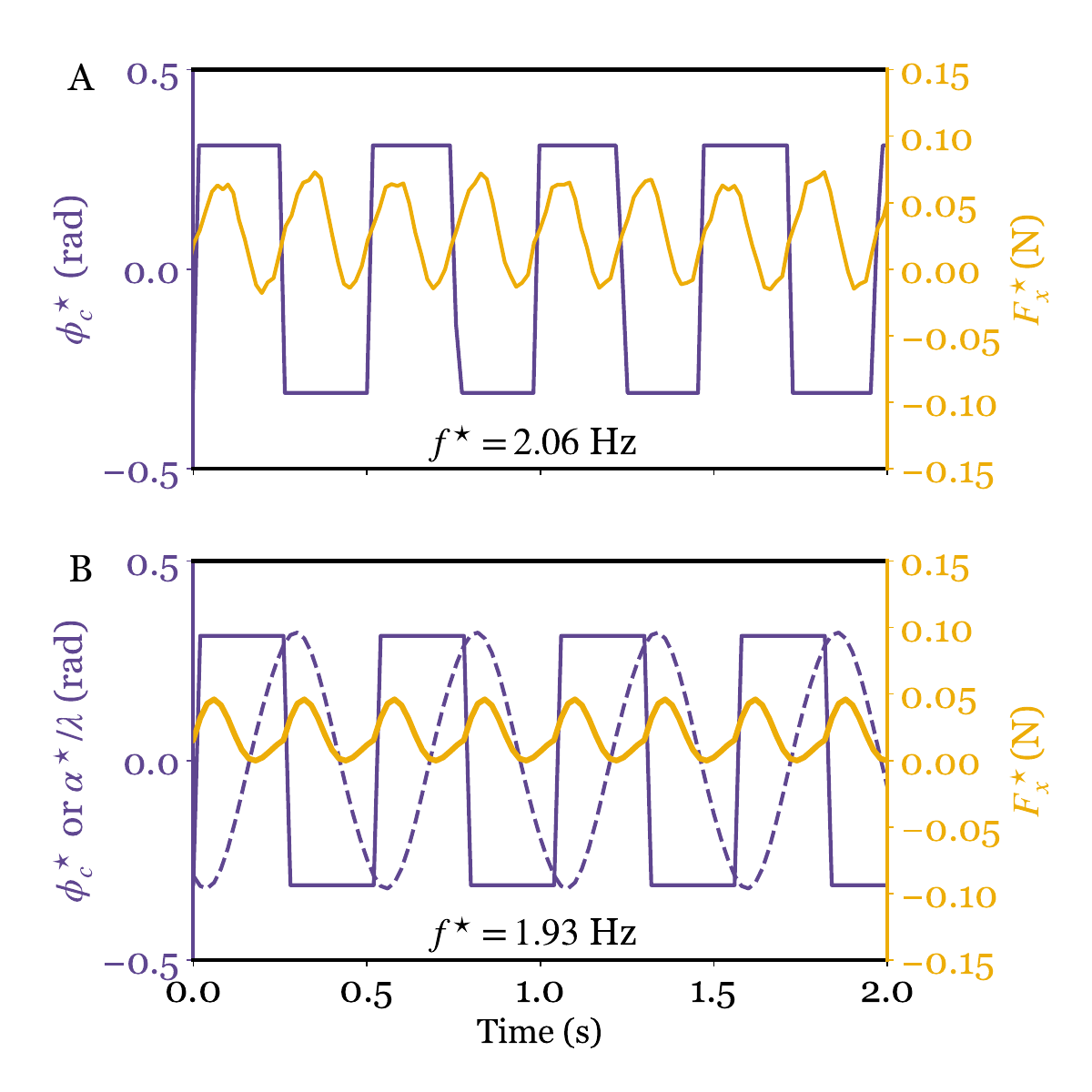}\\
        \caption{\label{fig:Inference}
         (A) Experimental result: instruction angle $\phi^\star_c(t)$ and $F^\star_x(t)$ of the best strategy determined by the algorithm, for $\Phi = 0.31\ \mathrm{rad}$.
         (B) Results from the model : temporal evolution of the instruction angle $\phi^\star_c(t)$ (purple solid line), the rescaled fin angle $\alpha^\star/\lambda$ (purple dashed line) and the resulting thrust $F^\star_x(t)$ (orange solid line), obtained for optimal control with the model and the parameters $\Omega=5.8$ rad.s$^{-1}$, $\Delta=0.29$ rad, $\omega_0=12.5$ rad.s$^{-1}$, $\xi=1.2$ and $\Phi = 0.31\ \mathrm{rad}$.
% \chr{Mettre dans caption la valeur de fréquence obtenue pour comparer avec fig. 5. Actualiser le texte et tourner positivement la petite différence en force. Parfait pour la forme de la commande. -> Li}
        }
    \end{figure}

First, the algorithm was provided with $\phi_c(t)$, $F_y(t)$ and its time derivative $\dot F_y(t)$ to account for the oscillatory nature of the system. After a few hours, the learning process converged toward a periodic motion. As shown in Fig. \ref{fig:Inference}A, the optimal control signal $\phi_c^\star(t)$ consists of a square wave function that switches abruptly between the two allowed extreme values~$\pm \Phi$.

Second, we provided images recorded by a camera, as shown in Fig. \ref{fig:setup}B, as inputs instead of relying on lateral force measurements. At each time step, four images were used (i.e., the current image and the previous three) to calculate convergence and speed. 
Despite this change in inputs, the resulting optimal control signal remained consistent with the previous findings. That is, the algorithm converged to a square wave function with the same amplitude and frequency, reaffirming the robustness of the previously identified optimal control strategy.

\begin{figure}[!b]
    \centering
\includegraphics[clip=true,width=\cl]{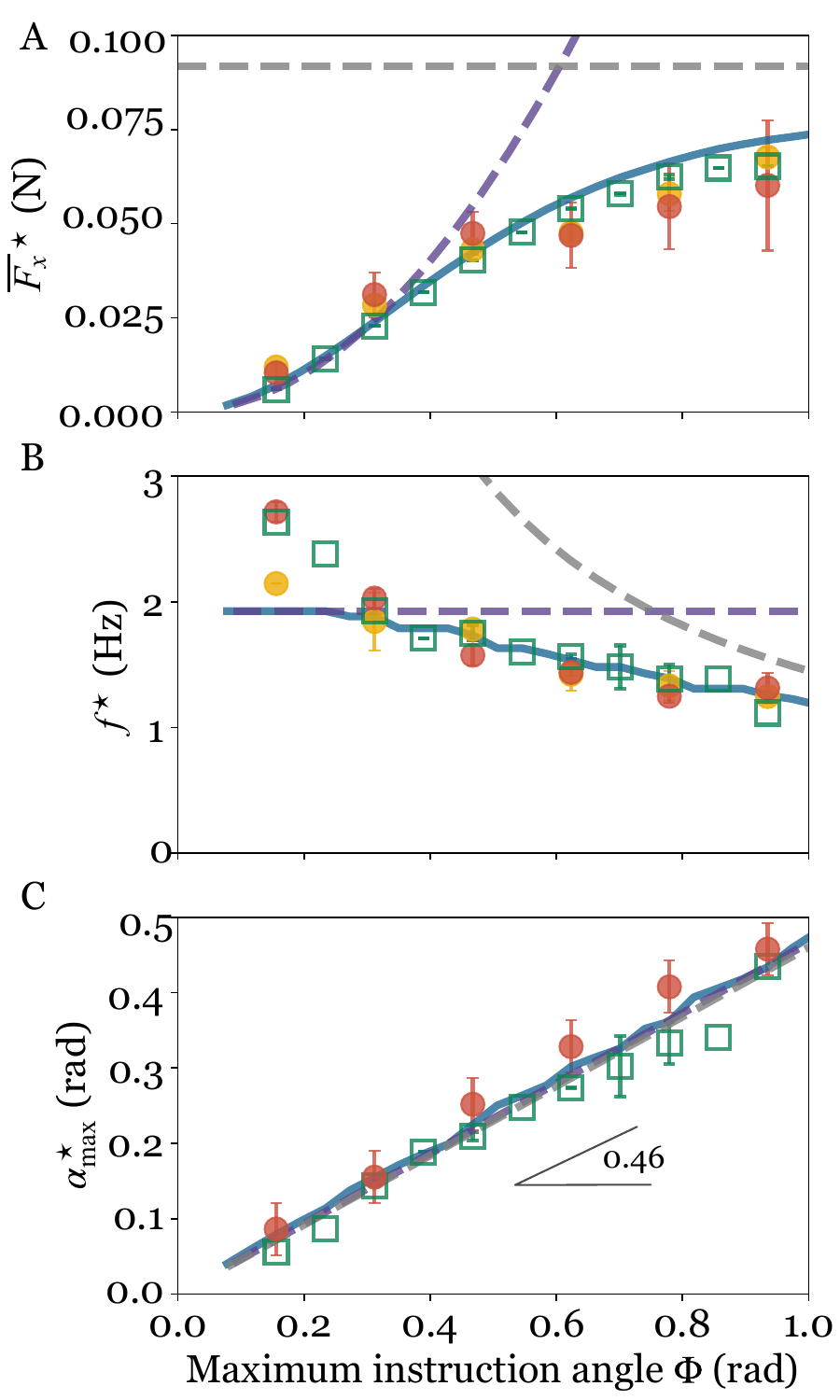}
 \caption{\label{fig:ResultsExpe}
 (A)
Maximum average thrust~$\overline {F_x^\star}$. 
(B) Optimal undulation frequency $f^\star$. 
(C) Maximal fin angle $\alpha^\star_\mathrm{max}$. 
For each figure, we compare the experimental results and those obtained with the model (Eqs. \ref{eq:mddtalpha}, \ref{eq:servomotor} and \ref{eq:averageThrust}). Experimental results: the red symbols are obtained via RL with the state $(F_y,\dot F_y,\phi_c)$; the orange symbols are obtained via image learning, where the state is a set of four successive pictures of the robot. Simulation results: the open square symbols denote RL results with the simulation of the model.
The dashed violet and gray curves represent theoretically predicted values in the limits of very small and very large $\Phi$, respectively. 
The solid green curve corresponds to the numerical determination of the frequency $f^\star$ that yields the highest thrust, $F_x^\star$, assuming a square waveform for $\phi_c$.
}
\end{figure}

The two approaches were then compared in terms of relevant physical quantities while varying the maximum instruction angle $\Phi$. In Fig.~\ref{fig:ResultsExpe}, the maximized average thrust $\overline{F_x}^\star$, the optimal frequency $f^\star$ and the corresponding maximum fin angle $\alpha^\star_\mathrm{max}$ are plotted as functions of $\Phi$. Whatever~$\Phi$, the agreement is excellent between the two approaches for all quantities. 
Specifically, $\alpha^\star_\mathrm{max}$ is proportional to $\Phi$ and the optimal frequency is a slowly decreasing function of $\Phi$ varying between $2.5$ and $1$~Hz in the parameter range considered.

\begin{figure}[!t]
    \centering
    \includegraphics[clip=true,width=\cl]{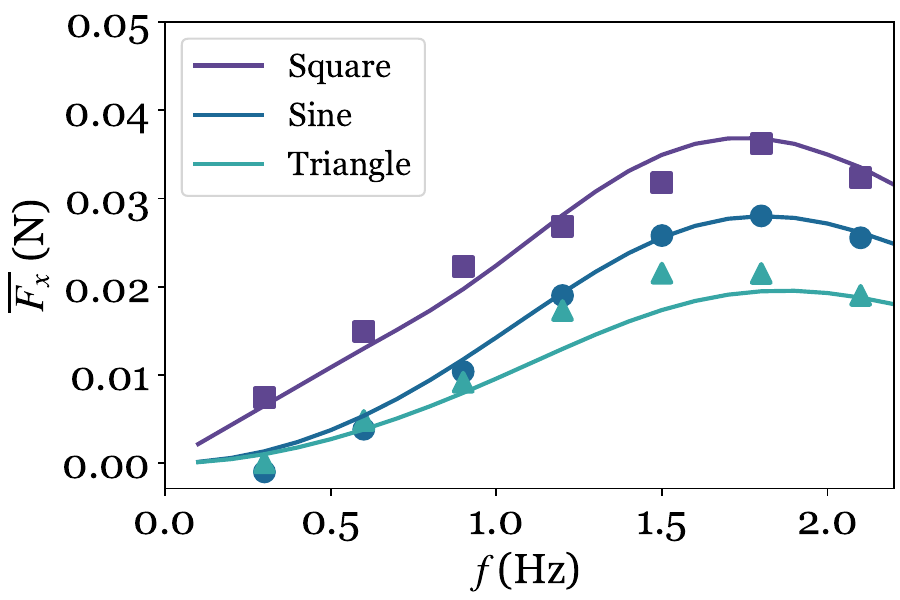}\\
    \caption{\label{fig:numWaveforms}
        Influence of the frequency on the thrust produced by some periodic functions:   
        square (purple squares), sinusoidal (blue circles) and triangular (green triangles) waves. Filled symbols and solid lines correspond to the experimental points and the model predictions, respectively. $\Phi = 0.42$ rad.  
    }
\end{figure}

To confirm that square wave forcing produces the greatest thrust, we applied three types of classical periodic forcing  (sinusoidal, triangular, and square) for a given $\Phi$ and measured the averaged thrust as a function of forcing frequency, $f$. Results are shown in Fig. \ref{fig:numWaveforms} for $\Phi = 0.42$ rad. Regardless of the type of actuation, the variation of the forcing frequency results in the presence of a peak in thrust, but the square wave forcing consistently generates the highest average thrust. This maximum thrust is achieved at a frequency of 1.8 Hz, a value that closely aligns with the frequency obtained using the RL approach. 
    
The analysis thus reveals a robust feature of optimal control: a square wave function oscillating between two extreme values.  This suggests the existence of a strong mechanism driving maximum thrust. 
\subsection{Model}
We present a comprehensive model that combines the physical aspects of underwater undulatory swimming with the internal dynamics of the servomotor to gain insights into the experimentally discovered optimal solutions.

The dynamics of the fin angle $\alpha(t)$ can be described by a damped harmonic oscillator, influenced by the angle $\phi(t)$ of the servomotor wheel \cite{sanchez-rodriguez_proprioceptive_2021}:
\begin{equation}\label{eq:mddtalpha}
    \ddot \alpha(t) +\xi \omega_0 \dot \alpha(t) +\omega_0^2(\alpha(t)-\alpha_c(t))=0, \ \alpha_c(t)=\lambda \phi(t) .
\end{equation}
This equation captures the interaction between the deformable fin and the surrounding water, driven by an instruction angle, $\alpha_c(t)$, that is proportional to the servomotor wheel angle, $\phi(t)$. 
The numerical values of the parameters were determined following the procedure described in~\cite{sanchez-rodriguez_proprioceptive_2021}.
The proportionality factor $\lambda=0.46$ depends on the length of the cables and elasticity of the polymeric skeleton.
Additionally, the servomotor has its own internal dynamics to adjust the servomotor wheel angle $\phi(t)$ to the instruction angle $\phi_c(t)$ \cite{sanchez-rodriguez_proprioceptive_2021}:
\begin{equation}\label{eq:servomotor}
    \dot \phi(t) =\Omega \tanh\left(\frac{1}{\Delta}\left(\phi_c(t)-\phi(t)\right)\right).
\end{equation}
The model includes the parameters $\Omega = 5.8$ rad.s$^{-1}$ and $\Delta = 0.29$ rad, which represent the maximum angular speed of the wheel and the required angle difference for the servomotor to operate at its maximum angular speed, respectively.
The nonlinearity introduced by the $\tanh$ function results in the saturation of the wheel speed $\dot \phi(t)$ at $\pm \Omega$ if the servomotor is unable to keep up with the provided instruction. This saturation occurs when the servomotor is too slow to reach the desired angle $\phi_c(t)$.

Eqs.~(\ref{eq:mddtalpha}) and (\ref{eq:servomotor}) correspond to two essential aspects of the dynamics: one pertains to the interaction between the undulating swimmer and its aquatic environment, while the other accounts for the limitations imposed by the swimmer's internal dynamics. Our experimental system thus serves as a suitable candidate for mimicking the dual nature of dynamics observed in natural fish. In natural fish, the internal dynamics are limited by biological constraints at the muscular level~\cite{sanchez2023scaling}.

The thrust is proportional to the mass of water accelerated in the longitudinal direction and is written $F_x=- K \alpha(t) \ddot \alpha(t)$~\cite{gazzola2015gait,sanchez-rodriguez_proprioceptive_2021}, where the parameter $K = 12.9 \ 10^{-3}$~N.rad$^{-2}$.s$^{2}$ characterizes the thrust efficiency of the robot in water and is measured following the procedure described in \cite{sanchez-rodriguez_proprioceptive_2021}. The average thrust over an undulation period $T=1/f$ can be written as:
\begin{equation}
          \overline {F_x} = - \frac{K}{T}\int_0^T \alpha(t) \ddot \alpha(t) dt  =  \frac{K}{T}\int_0^T \dot \alpha(t)^2 dt.
          \label{eq:averageThrust}
\end{equation}
We employ RL techniques on the model to verify the optimal control strategy that yields the maximum average thrust (see the Materials and Methods section for more details).
In Fig.~\ref{fig:Inference}B 
%\textcolor{red}{J: only to keep in mind for formatting. If they work the same way as natcomm they will ask to refer to Fig.2b before Fig.3}
, we show the time evolution of the optimal instruction angle $\phi_c^\star(t)$ for $\Phi = 0.31$~rad, alongside the corresponding rescaled fin angle $\alpha^\star(t)/\lambda$ and thrust $F^\star_x(t)$. 
Consistent with the experimental results, the optimal instruction angle adopts a square wave function, alternating between $\Phi$ and $-\Phi$. 

In fact, the dynamics presented in Figs.~\ref{fig:Inference}A and B, measured in experiments and simulated with the model, are very similar.  
This is confirmed by the excellent agreements shown in Figs.~\ref{fig:ResultsExpe}b and c for the relevant quantities such as undulation frequency and maximum fin angle of the optimal control. 
Furthermore, this simple model is able to reproduce the average thrust force, as depicted in Fig.~\ref{fig:ResultsExpe}A.

Regardless of the specific value of $\Phi$, both the experiments and the model demonstrate that the system waits for the fin angle to approach its instruction before changing the control. In other words, the control signal switches direction when the fin angle is close to its maximal value $\alpha^\star_\mathrm{max} = \lambda \Phi$. This finding suggests an intuitive mechanism for selecting the frequency simultaneously with the control switch, emphasizing the efficiency and adaptability of the system in achieving maximum thrust.

The model is also validated with the data obtained in Fig.~\ref{fig:numWaveforms} in which sinusoidal, triangular and square wave forms were enforced. The agreement between the experiments and the model is again excellent. The maximum thrust is generated around a frequency of $1.8$~Hz, indicating that the system operates optimally near its resonance in $\dot \alpha$, which occurs at a frequency $\omega_0/(2\pi) \sim 2.0$~Hz, regardless of the damping factor. For comparison, the resonance in $\alpha$ suggested by various other studies \cite{michelin_resonance_2009,paraz2016thrust,hoover2018swimming} occurs at a lower value, $\omega_0\sqrt{1-\xi^2/2}/(2\pi) \simeq1.1$~Hz in our system. 

\subsection{Bang-Bang control} 

The periodic and abrupt changes in the instruction angle resemble the behavior of a bang-bang controller \cite{sonneborn1964bang}.
In the context of the model, we apply Pontryagin's maximum principle \cite{pontryagin1987mathematical} to explain why square wave control at maximum allowed values $\pm \Phi$ in the instruction angle $\phi_c$ produces the maximum thrust $\overline {F_x}$, defined in Eq.~(\ref{eq:averageThrust}), under the constraints of Eqs.~(\ref{eq:mddtalpha}) and (\ref{eq:servomotor}).
    
The determination of the maximum thrust thus requires the resolution of a variational problem subject to three Lagrange multipliers, associated with the dynamic equations for $\alpha(t), \dot \alpha(t)$, and $\phi(t)$.
The bang-bang controller provides the optimal control if the response is driven by linear equations~\cite{kirk2004optimal}. This condition is not completely satisfied in this system due to the nonlinear behavior of the servomotor's internal dynamics, Eq.~(\ref{eq:servomotor}). However, a bang-bang controller remains the optimal choice in this system because of the specific nature of the nonlinear function driving the relaxation dynamics of $\phi$ (SI Appendix).

This appears particularly simple to explain for a fast servomotor, i.e., if $\phi(t)$ almost immediately follows the command $\phi_c(t)$. In our case, we can assimilate $\phi(t)$  to $\phi_c(t)$ for $\Phi\ll \Phi_s$, with:
\begin{equation}
    \Phi_s = \frac{\Omega}{\omega_0},
    \label{eq:phic}
\end{equation}
as shown in the SI Appendix. 
In both limits, the forcing command acts linearly through the Eq. (\ref{eq:mddtalpha}), which completely justifies the choice of a bang-bang controller as an optimal strategy.
In addition, this asymptotic permits the computation of the average thrust  resulting from a square wave driving at frequency $f$:
\begin{equation}
    \label{eq:fast_servo}
    \overline{F_x}= K \lambda^2\Phi^2 \omega_0^2   \frac{\frac{4 f}{\xi  \omega_0} \left( \sinh \left(\frac{\xi \omega _0}{4 f} \right)-\frac{\xi  \sinh \left(\sqrt{\xi ^2-4} \frac{\omega _0}{4 f}  \right)}{\sqrt{\xi ^2-4}}\right)}{\cosh
    \left(\sqrt{\xi ^2-4} \frac{\omega _0}{4 f} \right)+\cosh \left(\frac{\xi \omega _0}{4 f}\right)} ,
\end{equation}
where $K \lambda^2\Phi^2 \omega_0^2$ is the dimensional value that gives the scaling of the average thrust in this limit. Regardless of the value of $\xi$ between 0 and 2, this thrust is maximum when $\dot{\alpha}$ is resonant and the oscillator is driven close to its undamped  frequency:  $f^\star \simeq \omega_0/(2\pi)$.

For the case of a slow servomotor with  $\Phi \gg \Phi_s$, the system of equations is nonlinear, and the maximum thrust can be expressed as:
\begin{equation}
    \label{eq:slow_servo}
    \overline{F_x} = K \lambda^2 \Omega^2,
\end{equation}
which is reached as long as the servomotor wheel is moving at maximum angular speed $\Omega$ (i.e. $\dot{\alpha}(t)= \pm \lambda \Omega$ in Eq.~(\ref{eq:averageThrust})). The square wave forcing at maximum allowed values appears as the optimal solution to maximize the difference between $\phi_c(t)$ and $\phi(t)$ in Eq.~(\ref{eq:servomotor}) and ensures that the servomotor wheel is moving at maximum angular speed. 
In this limit, the optimal undulation period is determined by the shortest time required to sweep the servomotor angle from $-\Phi$ to $\Phi$ and then back to $-\Phi$, all at maximum angular speed. This leads to the expression $f^\star = \Omega / (4 \Phi)$.

The behavior resulting from both limits is depicted in Fig.~\ref{fig:ResultsExpe}, and agrees with the outcomes obtained in experiments and with the model, as well as indicating a transition for $\Phi$ around $\Phi_s$.

\subsection{Swinging control: a model-free strategy}

In the above model, to maximize thrust, the optimal frequency must be found in order to select the best square wave function for control. It is therefore necessary to perform preliminary calibrations to measure the relevant quantities, in this case $\omega_0$, $\xi$, $\Omega$ and $\Delta$, to bring a particular system closer to optimality. Here we explore another strategy that does not require any prior knowledge about the system.

Starting from the thrust expression, Eq.~(\ref{eq:averageThrust}), the optimization is intrinsically linked to the time evolution of the fin angle velocity $\dot{\alpha}$. An intuitive approach would be to favor the phases of maximum speed by reversing the sign of the instruction angle when the fin slows down too much. We call this swinging control in reference to the way children are able to swing without knowing the physical laws involved.

To this end, we introduced an instruction-changing criterion $C$ ($0\leq C \leq 1$), such that $\phi_c$ changes sign if the speed slows to $C\dot{\alpha}_M$, with $\dot{\alpha}_M$ the last observed maximal angle speed. This condition ensures that the highest possible speed is maintained and that the instruction is changed when the speed becomes too slow. We found numerically that this strategy is always independent of the system history and the initial conditions, as the algorithm converges toward a unique solution.

\begin{figure}[htb]
       \centering
    \includegraphics[clip=true,width=\cl]{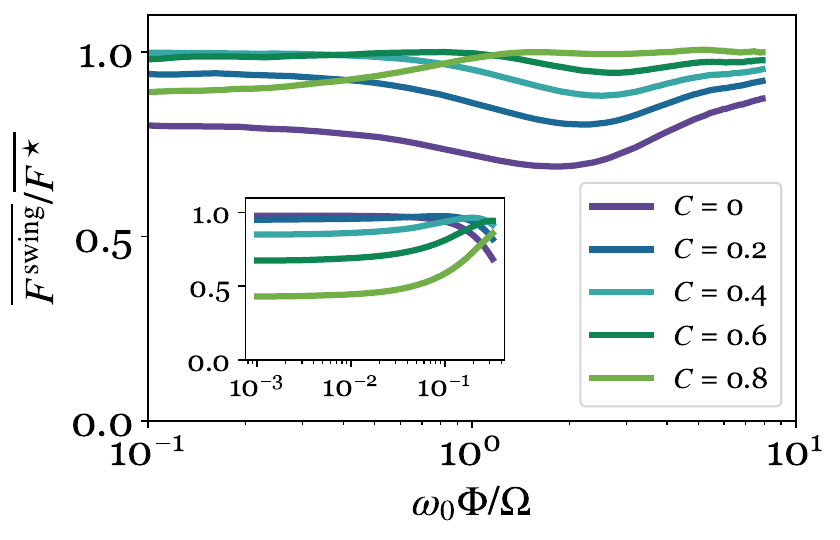}\\
        \caption{
        Performance of the swinging strategies compared to the optimal solutions as a function of $ \Phi/\Phi_s=\omega_0 \Phi/\Omega $, with different instruction-changing criteria. We fixed $\omega_0=12.5$~rad.s$^{-1}$, $\xi=1.2$, $\Omega = 5.8$~rad.s$^{-1}$ and $\Delta=0.29$~rad.
        Inset: Performance of the swinging strategies with the same instruction-changing criteria for a fast servomotor $\omega_0 \Phi/\Omega <1$  and $\xi = 0.05$.
    }
    \label{fig:comparaisonSwing}
    \end{figure}

We evaluated the average thrust resulting from the swinging strategy by varying the parameters $\omega_0 \Phi / \Omega$, which measures the servomotor's ability to follow the instruction, and $C$. 
Figure~\ref{fig:comparaisonSwing} shows that a criterion value of $C$ ranging from 0 to 0.8 yields satisfactory performance for the swinging strategy whatever the value of $\omega_0 \Phi / \Omega$. The swinging control consistently delivers excellent outcomes without imposing stringent constraints on the choice of $C$:  the thrust efficiency, defined as $\overline{F^\mathrm{swing}}/\overline{F^\star}$,  being higher than 70\% in the parameter range: for  $C=0.6$, this ratio exceeded 95\% regardless of the value of $\omega_0 \Phi / \Omega$, but fine-tuning remains possible. 
This strategy can be proposed even in specific cases : 
in the example of a fast servomotor ($\Phi  \ll \Phi_s$) and an undamped oscillator ($\xi \rightarrow 0$), the optimum is attained when the instruction angle changes sign at $\dot{\alpha} = 0$ (i.e., $C=0$) with a thrust efficiency very close to 100\%, as shown in the inset of Fig. \ref{fig:comparaisonSwing}, and detailed in (SI Appendix).

Swinging control is a robust strategy for achieving the highest thrusts without prior knowledge of the system or complex control algorithms. Its straightforward implementation using basic sensors makes it practical for various applications.

\subsection{2D direct numerical simulations}

To transpose our results to a real swimming problem, we conducted simulations for the complete fluid-structure interaction in a 2D configuration (see details in the Materials and Methods section).
In this numerical setup, the swimmer moves within a tank filled with liquid having the same density and viscosity as water.
We consider the fish body to be viscoelastic, and we adjust its parameters to match the values of $\omega_0$ and $\xi$ obtained from the robotic fish experiments.  The entire body measures $L=10\ \mathrm{cm}$ with an average thickness equal to $H=1.3\ \mathrm{cm}$. 
To model muscular activity, we impose a spatiotemporal variation of the elastic body length at equilibrium.
In particular, we  set that the equilibrium of the strain component $\epsilon_{xx}$ varies parabolically as a function of the distance to the head $X$, and linearly from the midline $Y$: $\epsilon_{xx}\propto (X/L)^2 (Y/H)a(t)$, where $a(t)$ drives the swimmer deformation and varies temporally. 
With this functional form, the half body ($Y>0$) extends its length while the other part ($Y<0$) retracts, resulting in the swimmer bending to compensate for the inhomogeneous change in length across the body thickness. 
We have simulated the motion of this 2D active elastic beam embedded in water, driven by different functional forms of $a(t) \in (-1,1)$.  

\begin{figure}[thb]
        \centering
        \includegraphics[clip=true,width=\cl]{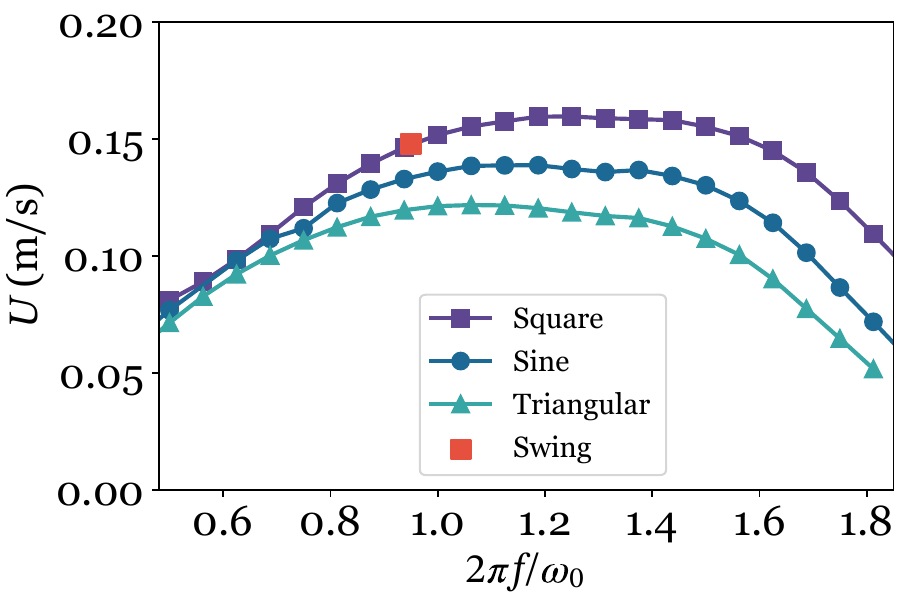}
        \caption{\label{fig:comsolRamp}
        Computation of the swimming speed using a full 2D Fluid-Structure Interaction simulation.
        The solid lines and symbols represent the forcing with different wave shapes and the red square corresponds to the swinging controller.
    }
    \end{figure}

In Fig.~\ref{fig:comsolRamp}, we present the cruising speed achieved with square, sine, and triangular wave functions at various control frequencies~($f$). The square wave forcing consistently leads to the highest speed, regardless of the frequency, confirming the predictions from the RL algorithm and the model.
In addition, we have implemented the swinging strategy with $C=0$.  The swinging controller automatically selects a frequency that propels the swimmer to a speed close to the maximum, as depicted in Fig. \ref{fig:comsolRamp}. Therefore, our numerical simulations validate our interpretations regarding the mechanisms to achieve the highest swimming speed.  
    
\section{Discussion}
In this article, our primary objective was to determine the actuation that yields the highest thrust, and thus the highest swimming speed, for underwater swimmers. Considering the complex fluid-structure interaction at relatively high Reynolds numbers, we employed various approaches and methods to shed light on this topic. Notably, reinforcement learning techniques played a crucial role in revealing that the most effective method to achieve high thrust in experiments involves employing a bang-bang controller. This type of actuation oscillates abruptly and periodically between the two extreme command values.

To rationalize and validate these experimental results, we studied a simple model using reinforcement learning. The results from this model closely aligned with those obtained through the experimental approach, further reinforcing the effectiveness of the bang-bang controller. Additionally, full 2D numerical simulations of autonomous swimmers confirmed the validity of our findings.

We have successfully illuminated the strategy to attain the highest thrust for underwater swimmers, comprising three key messages. Firstly, the command should resemble a periodic square function, exploiting the natural oscillation of the tail-fin system. Secondly, the frequency of actuation needs to be close to that which maximizes tail speed. This frequency tends to approach the body's natural undulation frequency as damping vanishes. 
Lastly, an efficient approach to selecting a nearly optimal swimming gait is to switch the actuation sign as the fin speed becomes too small.

Given the simplicity and practicality of our results, we envision powerful applications in terms of underwater biomimetic swimmers. Furthermore, these findings can be utilized to develop strategies that enhance the performance of human athletes, particularly those involved in aquatic disciplines. The potential for real-world implementation is promising, opening up exciting avenues for both biomimetic technology and athletic performance improvement.

\begin{acknowledgments}
We thank F. Boyer for very interesting comments.\\
This work has been supported by UCAJEDI, UCA NIDS, with the reference  ANR-15-IDEX-0001 , UCA DS4H,with the reference ANR-17-EURE-0004 and Ministerio de Universidades and European Union-NextGenerationEU.
\end{acknowledgments}

\appendix

\section{Experimental setup}

To measure forces, a force sensor is linked to the robot fish via an aluminum rod, enabling bi-directional force measurements (longitudinal force, $F_x$, and normal force, $F_y$) with a precision of approximately $10^{-3}$ N. An analog force signal is converted into a digital format using an ADC converter (Adafruit 1115) and then collected by the Raspberry Pi via the I2C interface.

The robotic swimmer is positioned in a water tunnel (Rolling Hills Research Corporation, Model 0710) using a beam clamp that holds the aluminum rod. In this experimental study, our focus is on determining the highest thrust generated by the oscillation of the swimmer's tail; hence, no water flow was generated by the tunnel pump.

For the learning process based solely on force sensor measurements, we conducted it directly on the Raspberry Pi. However, for the image-based learning process, we utilized an external computer due to computational constraints. Data exchange between the Raspberry Pi and the computer occurred via the TCP/IP protocol using an RJ45 Ethernet cable.

\section{Machine Learning with force sensor\label{sec:Machine Learning}}
%\subsection{}
In our search for the best functional form for the servomotor driving, we employed deep Reinforcement Learning (RL) techniques. Within this framework, we described the control of the swimmer using Markov decision processes (MDP).

The state of the swimmer, denoted by $s=(F_y(t),\dot{F_y}(t),\phi_c(t))$, 
%\textcolor{red}{state is $\phi_c$ or $\phi$? In caption of Fig.3 $\phi$ is written} \textcolor{blue}{Li: good remark, corrected in the caption}
 is characterized by the normal force, $F_y(t)$, and its derivative, $\dot{F_y}(t)$, as well as the command angle of the servomotor, $\phi_c(t)$. The action, represented by $\phi_c(t) \in [-\Phi,\Phi]$, refers to the command angle of the servomotor, and the reward is directly proportional to the thrust force, $F_x(t)$. The objective of the RL algorithm is to maximize the total cumulative reward, which, in this context, translates to maximizing the thrust, $F_x$, generated by the swimmer.

To achieve this, the RL algorithm explores the available action and state spaces and gradually converges to the best control sequence through a trial and error process. We exploit the PPO (proximal policy optimization) algorithm~\cite{Schulman2017} for this study because it is capable of handling both discretized and continuous action spaces. PPO is a policy gradient method in RL that stabilizes the learning process to avoid large excursions.

In our experiments, RL applies the control policy and sends a command to the robotic fish every $50$~ms. The control policy is incrementally updated (trained) after each $768$ control steps. We chose this number to ensure sufficient exploration of the action and state spaces before each update. The entire training process spans $10^5$ control steps, which amounts to approximately $2$ hours. Throughout this process, we save the control policy and the state value function every $5000$ steps.

After each training of the Neural Network, we conduct an inference to evaluate the quality of the actual best control policy:  only the best action is chosen at each step during this process.  This evaluation provides valuable insights into the performance of the swimmer with the optimized control sequence. More details regarding the PPO implementation and the meta-parameters are given in the Supplementary Text file.

\section{Machine learning with images}
In our study, we employed a web-camera (ODROID USB-CAM 720P) placed above the robotic fish's undulating fin to record its motion. Instead of the previous state representation $(F_y(t),\dot{F_y}(t),\phi_c(t))$, we used a stack of 4 consecutive images separated by the sampling time interval \cite{mnih2013playing}, to represent the state, $s$, of the robotic fish. 
The recorded RGB images were preprocessed by converting them to gray-scale and re-scaling them from their original size to $(84, 84)$ pixels. 

This resizing was performed to reduce the computation time, while maintaining sufficient information. We used a Convolutional Neural Network (CNN), as introduced by \cite{lecun1989handwritten,lecun2015deep}, as an image compression tool for the raw data of the images ($84 \times 84 \times 4$~pixels). The output of the CNN is a latent vector with $512$ dimensions that is then passed to the fully-connected neural networks to perform the actor-critic algorithm.
Beyond this change in state representation, the remaining experimental methods for learning remained the same as those used for learning from the force sensor.
    
\section{Fluid-Structure Interaction Simulation\label{sec:Comsol Simulation}}
To simulate the fish swimming driven by an optimal command, we used the software COMSOL~6.1 following the approach outlined in  \cite{curatolo2015virtual}. The 2D computational domain covers a rectangle with dimensions $100 \times 20$ cm², representing water.

\begin{figure}[thb]
        \centering
        \includegraphics[clip=true,width=\cl]{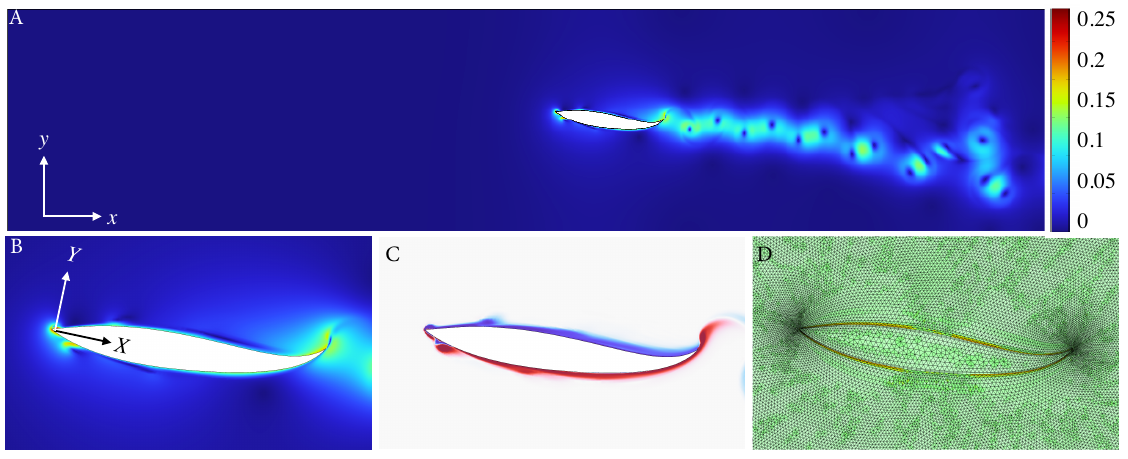}
        \caption{\label{fig:setupComsol}
        (A) Setup of the computational domain. 
        The white region represents the swimmer body, and
        the color codes the speed in m/s.
        (B) Zoom around the swimmer body.
        (C) Typical vorticity field around the swimmer.
        (D) Zoom around the swimmer to show the typical mesh precision used for the simulations.
        }
    \end{figure}

In Fig.~\ref{fig:setupComsol}A, we show the geometry of the simulation.
The two horizontal borders represent slip boundary conditions for the velocity, while the left and right borders are associated with entrance and exit boundary conditions. The swimmer is approximated by a viscoelastic beam of Young Modulus $10^4 \ \mathrm{Pa}$, a Poisson ratio $0.3$ and a viscosity $10^{-4}\ \mathrm{Pa.s}$. 
It navigates toward the left. The thickness~$t(X)$ of the viscoelastic beam is given by the relation
$$
    t(X)=H \frac{X}{L}\left(1-\frac{X}{L}\right)e^{-\frac{X}{L}} ,
$$
where $X$ is the curvilinear distance along the midline, measured from the head. Here $H=4\ \mathrm{cm}$ and $L=10\ \mathrm{cm}$, such that the thickest part of the beam measures $1.3\ \mathrm{cm}$, see Fig.~\ref{fig:setupComsol}B.
To model the antagonistic muscle action, we impose on the swimmer that the equilibrium component $\epsilon_{xx}$ of the strain varies spatiotemporally:
$$
\epsilon_{XX}(X,Y,t)=0.01(X/L)^2 (Y/H)a(t) ,
$$
where $a(t)$ drives the  motion dynamics, and $\epsilon$ is the strain tensor~\cite{landau1986Elasticity}. This forcing modifies the equilibrium length of each part of the body in an opposite manner: when the superior part of the swimmer elongates its length, the other part contracts it, following the dynamics of $a(t)$. $a(t)$ can be a wave function or defined as $a(t)=\mathrm{sign}(v(t))$, where $v(t)$ is the normal velocity of the swimmer at the tail.
The numerical value is chosen such that the typical amplitude of the tail oscillation is close to 0.2.

The fluid-structure problem is solved using a fully coupled approach and the PARDISO linear solver; the nonlinear problem is tackled with a Newton algorithm. Because the swimmer deforms its shape, the mesh is adapted using a Yeoh method. To avoid excessively large deformations in the mesh due to the swimmer's movement, the entire computational domain is remeshed is the discretisation is too distorted. Approximately 7,000 vertices are needed for almost 40,000 degrees of freedom to coarsely solve the complete fluid-structure interaction on the whole domain, and predict a correct swimming velocity.
A finer element distribution is ensured at the head and tail (Fig.~\ref{fig:setupComsol}D.)
  However, to accurately capture the wake, 300,000 elements are required, as shown in Fig.~\ref{fig:setupComsol}D. The typical time step used is $10^{-3}$ s. The center of mass of the swimmer is computed at each time step during the simulation. This comprehensive setup enables us to study and analyze the swimming behavior of the robotic fish driven by the optimal command obtained through the reinforcement learning process.

\bibliographystyle{Science}

\clearpage

\end{document}

% Include your paper's title here

\title{Supplemental Material\\
\vspace{1cm}
Optimum control strategies for maximum thrust production in underwater undulatory swimming}

% Place the author information here.  Please hand-code the contact
% information and notecalls; do *not* use \footnote commands.  Let the
% author contact information appear immediately below the author names
% as shown.  We would also prefer that you don't change the type-size
% settings shown here.

\author{Li Fu}
 \affiliation{Laboratoire de Tribologie et Dynamique des Systemes, \'Ecole Centrale
de Lyon, CNRS}%Lines break automatically or can be forced with \\
\author{Sardor Israilov}
 \affiliation{Universit\'e C\^ote d'Azur, CNRS, INPHYNI, 17 rue Julien Laupr\^etre, 06200 Nice, France}
\author{Jes\'us S\'anchez-Rodr\'iguez}%
\affiliation{Departamento de F\'isica Fundamental, Universidad Nacional de Educaci\'on a Distancia,  Madrid, 28040, Spain}

\author{Christophe Brouzet}
 \affiliation{Universit\'e C\^ote d'Azur, CNRS, INPHYNI, 17 rue Julien Laupr\^etre, 06200 Nice, France}

 \email{Second.Author@institution.edu}

\author{Guillaume Allibert}
 \affiliation{Universit\'e C\^ote d'Azur, CNRS, I3S, Sophia Antipolis, France}

\author{Christophe Raufaste}
 \affiliation{Universit\'e C\^ote d'Azur, CNRS, INPHYNI, 17 rue Julien Laupr\^etre, 06200 Nice, France}
 \affiliation{IUF, Paris, France}
\author{M\'ed\'eric  Argentina}
 \affiliation{Universit\'e C\^ote d'Azur, CNRS, INPHYNI, 17 rue Julien Laupr\^etre, 06200 Nice, France}
 
\date{\today}% It is always \today, today,

% Double-space the manuscript.

%\baselineskip24pt

% Make the title.

\maketitle 
\onecolumngrid
\section*{Reinforcement Learning}
%\subsection{Preliminaries}
Reinforcement learning (RL) has achieved remarkable success in controlling dynamic systems, both in simulations and real-world applications. 
RL, a subfield of machine learning, focuses on how agents make decisions within an environment to maximize cumulative rewards. 
Here, an agent refers to an algorithm that interacts with the environment and takes actions at each time step. 
RL leverages the framework of Markov Decision Process (MDP), which defines sets of states $s\in S$, actions $a \in A$, transition probabilities $P(s,a,s')$ (between states $s$ and $s'$ under the action $a$) and rewards $R(s,a)$ (for taking an action $a$, while in the state $s$).
In accordance with the MDP, the state at time  $t+1$ $s_{t+1} $ depends on what occurs at time $t$$ : s_t, a_t $, and the transition probability $ p(s_t,a_t,s_{t+1}) $. In other words, the agent's state can change based on the evolution of the environment and its actions. The agent operates according to its own strategy named policy $ \pi(s,a) $, which is the probability of taking an action $a$ while in state $s$. After each learning iteration, the updated state and the associated reward will be informed to the agent and the latter will update its policy in order to maximize the cumulative reward.

In our experiments, the environment involves a bio-mimetic fish interacting with water.
States $s\in S$ correspond to $s=(F_y,\dot{F_y},\phi_c)$, actions $A$ to $[-\Phi,\Phi]$ and rewards to the the thrust force $ F_x $. The RL objective is to maximize the discounted expected reward $ F_x $: $\mathbb{E}^\pi\left[\sum_{t=0}^{+\infty} \Omega^t R\left(s_t, a_t\right)\right]$ with $\Omega$ a discount factor, by optimizing its action policy $\pi(s,a)$.

During the training process, the agent interacts with its environment through a series of episodes, each encompassing a fixed span of 768 time steps in our case. This episodic structure fosters a more stable and consistent convergence by mitigating the potential accumulation of errors. 
Following the completion of each episode, a designated pause is instituted to allow sufficient time for the variables $F_x$, $F_y$, and $\dot{F_y}$ to revert to a baseline value of zero. Concurrently, the parameter $ \phi_c $ is reset to zero, thereby enhancing the reliability and accuracy of the learning process.

We recall the definitions of the value function $V(s) = \underset{a \sim \pi}{\mathbb{E}^\pi}\left[\sum_{t=0}^{+\infty} \Omega^t R\left(s_t, a_t\right)\right | S_0=s]$, which is the expectation of the total cumulative reward under the current policy $\pi(s,a)$ of being in a state $s$; and the value-action function (Q-value) $Q(s,a) = \underset{a \sim \pi}{\mathbb{E}^\pi}\left[\sum_{t=0}^{+\infty} \Omega^t R\left(s_t, a_t | S_0=s, A_0=a\right)\right]$, which is the expectation of the total cumulative reward after taking action $ a $ at the initial state $ s $. 

\subsection*{PPO}
%We use the Proximal Policy Optimization (PPO) algorithm \cite{Schulman2017} in this study. 
%It is an actor-critic algorithm \cite{Ruckstiess2008} with the advantages normalization technique. The code is implemented in the framework of the open-source library Stable Baseline 3 \cite{stable-baselines3}. The pseudo-code of the algorithm (PPO) is shown in Algorithm \ref{algo:ppo}. 

%PPO has two neural networks (NN) named actor NN and critic NN. The Actor NN represents the policy $ \pi_{\theta}(s,a) $ and is updated according to a variant of policy gradient with normalized advantage estimation \textbf{$ \hat{A}_t $}. The advantage estimation measures how good the taken action is with respect to the predicted one. PPO uses techniques such as gradient clipping and KL-divergence computation between 2 consecutive policies to stabilize the training. The critic NN evaluates the value function after the agent takes an action according to the actor NN. 

In this study, we use the Proximal Policy Optimization (PPO) algorithm [43]. Functioning on the backbone of an actor-critic framework [49], PPO leverages advanced normalization techniques to optimize the learning process. Our implementation of the algorithm is based on the Stable Baselines 3 library [50]. The pseudo-code of the algorithm (PPO) is illustrated in Algorithm \ref{algo:ppo}. 

Central to the PPO algorithm are two intertwined neural networks: the actor and the critic, each fulfilling distinct yet complementary roles. On one hand, the actor neural network, denoted as actor NN, embodies the policy function ($ \pi_{\theta}(s,a) $), which undergoes periodic updates through a refined policy gradient method that incorporates normalized advantage estimation \textbf{$ \hat{A}_t $}. This estimation serves as a metric to gauge $Q(s,a)$, the return of the chosen action in comparison to the expected return of the current state $V(s)$.
To ensure a stable and robust training phase, PPO employs the so-called clipped surrogate objective function and incorporates the entropy bonus, enabling a balance between exploration and exploitation during the learning process. On the other hand, the critic NN operates as an evaluator, assessing continuously the value function, and providing feedback for policy optimization.

The objective function is formulated as :

\begin{equation}
    L_t(\theta, \omega)=L_\mathrm{policy}(\theta) - c_{v}*L_\mathrm{value} + c_{e}*L_\mathrm{entropy}, 
\end{equation}
where $L_\mathrm{policy}(\theta)$ is the policy surrogate, which should be maximized, but PPO restricts the update to be within a small range defined by $\epsilon$ to prevent too large policy updates, which improves the stability in training. $L_\mathrm{value}(\omega)$ is the value function loss, measuring the difference between the value function estimated by the current critic NN and the expected value $\hat{R}_{t}$. Finally, incorporating the entropy bonus $L_\mathrm{entropy}$ encourages exploration by discouraging premature convergence to a deterministic policy. $c_v$ and $c_e$ are two coefficients to weigh the value function loss and the entropy bonus, respectively. These are hyperparameters that need to be tuned.

The three objective/loss functions are formulated as follows:
    \begin{eqnarray}
    L_\mathrm{policy}(\theta)&=&\hat{\mathbb{E}}_t \left[ \min \left(r_t(\theta) \hat{A}_t, \operatorname{clip}\left(r_t(\theta), 1-\epsilon, 1+\epsilon\right)\right) \hat{A}_t \right], \\
    L_\mathrm{value}(\omega)&=& \hat{\mathbb{E}}_t \left[(V_{\omega}(s_t)-\hat{R}_{t})^{2}\right],\\
    L_\mathrm{entropy}&=& - \hat{\mathbb{E}}_t \left[ \pi(s_t, a_t) \log \pi(s_t, a_t) \right],
\end{eqnarray}
where $r_t(\theta)$ is the probability ratio, defined as $\frac{\pi_\theta(s_t, a_t)}{\pi_\mathrm{old}(s_t, a_t)}$, representing the probability of taking action $a_t$ at state $s_t$ in the current policy compared to the previous one. 

\begin{algorithm}[H]
	\caption{Proximal Policy Optimization (PPO-Clip) Actor-Critic Style }\label{algo:ppo}
	\begin{algorithmic}[1]
		\State Input: initial policy parameters $\theta_0$, initial value function parameters $\omega_0$, clipping threshold $\epsilon$.
		\While{time step $< 10^5$} 
		\State Run policy $ \pi_{\mathrm{old}} $
		in environment for T time-steps and put data in the buffer
		\For{epoch = 0, 1, 2, ... K} 
		\State Take a batch size of trajectories $ (s_t,a_t,s_{t+1})$ in the buffer and
		compute advantage estimates $ \hat{A}_t $
		\State  Optimize the loss Function $L_t(\theta, \omega)$ with respect to  $ \theta $ and $\omega$ , with  T  gradients using minibatch size M
		\State Update $\theta$ and $\omega$
		%\State $ \theta_{old } \leftarrow \theta $
% 		\State 
		\EndFor
		\EndWhile
	\end{algorithmic}
\end{algorithm}

\subsection*{Hyperparameters}
The parameters were tuned with the help of Optuna framework [51]. The complete list of hyperparameters are:
    \begin{itemize}
	\item \textbf{Numbers of epochs \textbf{K}}: number of passes that PPO learns the buffer data.
	\item \textbf{Batch size \textbf{B}}: number of samples used for the gradient descent update of neural network. In practice, it should be large enough to avoid biased experience, but not too large to slow the learning. Furthermore, in policy gradient updates, a small batch size may destabilize the learning due to large variance.
	\item \textbf{Learning rate $\alpha $}: an extent to which we update the neural networks at each step.
	\item \textbf{Discount factor $\gamma$}: discount rate of future steps, which tells how the present is more valuable than the future.
	\item \textbf{Network size $N_s$}: the size of the dense neural network for "value" and "policy" function approximations.
	\item \textbf{Number of gradient steps T}: how many gradient steps of learning do we make during every epoch.
	\item \textbf{Entropy coefficient $ c_e $}: the weight of entropy of actions in the total optimized loss. Entropy signifies how randomly the taken actions are. It permits to explore the unknown states.
	\item \textbf{Value coefficient $ c_v $}: the weight of value regression in the total optimized loss.
 
 The CNN architecture used in the algorithm is the same as in [52].
 
\end{itemize}
\begin{table}[htb]
	\centering
	\begin{tabular}{p{6cm}p{4cm}}
		Name & Value\\
		\midrule
            Number of epochs $K$ & 40 \\
		discount factor (\textbf{$ \gamma $}) & 0.98\\
		epsilon (\textbf{$ \epsilon $}) & 0.3\\
		Batch size (B) & 128 \\
            Network size (\textbf{$N_s$}) & 2 layers with 64 nodes \\
		Learning rate (\textbf{$ \alpha $}) & $ 1.7. \space 10^{-5} $\\
		%$ \lambda_{\Omega} $ (C) & 1.0 \\ \hline
		
		Gradient steps (T) & 768\\
		Value coefficient ($ c_v $) & 0.451 \\
		Entropy coefficient (\textbf{$ c_e $}) & 0.0876\\
		\midrule
	\end{tabular}
	\caption{\label{TabHyperparameters} PPO hyperparameters}
\end{table}

\section*{Why a bang-bang controller is optimal?}
Following the Pontryagin principle, we aim to determine the best command to optimize the dimensionless force, averaged for a duration $T$ :
\begin{equation}
    I=\frac{\overline{F_x}}{K \Lambda^2\omega_0^2}=\frac{1}{\Lambda^2\omega_0^2}\frac{1}{T}\int_0^T \dot\alpha ^2 dt, \quad \Lambda=\lambda\Phi
\end{equation}
under the dynamics :
\begin{eqnarray}
    \label{eq:dotalpha}
    \dot \alpha&=&\omega\\
    \label{eq:dotomega}
    \dot \omega &=& -\xi\omega_0 \omega-\omega_0^2(\alpha-\alpha_c), \ |\alpha_c|\leq\Lambda, \quad \alpha_c=\lambda \phi\\
    \label{eq:dotphi}
    \dot \phi&=&\Omega \tanh\frac{\phi_c-\phi}{\Delta}.
\end{eqnarray}
Within the variational methods framework,  we write the Lagrangian:
\begin{equation}
    L=\omega^2+p_1(\dot \alpha-\omega)+p_2(\dot\omega+\xi\omega_0\omega+\omega_0^2(\alpha-\lambda \phi))+p_3\left(\dot\phi-\Omega \tanh\frac{\phi_c-\phi}{\Delta}\right),
\end{equation}
where we have introduced three Lagrangian multipliers, $p_{i=1..3}$ to impose the dynamics of $\alpha$, $\omega$ and $\phi$.
We reformulate this problem using the natural Hamiltonian:
\begin{align}
    H&=-L+p_1\dot\alpha+p_2\dot\omega+p_3\dot \phi\\
    H&=-\omega^2+p_1 \omega+p_2\left(-\xi\omega_0 \omega-\omega_0^2(\alpha-\lambda \phi)\right)+p_3 \Omega\tanh\frac{\phi_c-\phi}{\Delta}.
    \label{eq:Hamiltonian}
\end{align}

The co-state equations for this system are :
\begin{align}
    \dot p_1&=-\partial_\alpha H=\omega_0^2 p_2 \\
    \dot p_2&=-\partial_\omega H=2\omega-p_1+\xi \omega_0 p_2 \\
    \dot p_3&=-\partial_\phi H=-\omega_0^2\lambda p_2 +p_3 \frac{\Omega}{\Delta}\cosh^{-2}\frac{\phi_c-\phi}{\Delta}
\end{align}
Following the Pontryagin seminal idea, the value of the control parameter $\phi_c$ can be chosen in order to maximize the Hamiltonian value.
The bang-bang controller will be optimal if the fourth term in the Hamiltonian (\ref{eq:Hamiltonian}) which contains $\phi_c$ is linear, which is not the case here.
Nevertheless, a square forcing for $\phi_c=\pm \Phi$ will maximize the Hamiltonian (\ref{eq:Hamiltonian}): if $p_3>0$, we choose $\phi_c=\Phi$ and $\phi_c-\phi$ will be positive. On the contrary, if $p_3<0$, we choose $\phi_c=-\Phi$ and $\phi_c-\phi$ will be negative:
\begin{equation}
    \phi_c=\Phi\ \mathrm{sign}(p_3),
\end{equation}
and we recover a bang-bang controller.

\section*{Conditions for ensuring that a servomotor will perfectly follow the consign}
In this section, we perform a dimension analysis to evaluate the rapidity of the servomotor.
We make the following change of variables:
\begin{align}
s&=\omega_0 t\\
\alpha(t)&=\lambda\Phi \beta(s)\\
\phi(t)&=\Phi \psi(s),
\end{align}
and we get the system to solve:
\begin{align}
    \ddot \beta(s)&+\xi \dot \beta(s)+\beta(s)-\psi(s)=0\\
    \dot \psi(s) &=\frac{\Omega}{\Phi \omega_0}\tanh\left(\frac{\Phi}{\Delta}(\psi_c(s)-\psi(s))\right),
\end{align}
which uncovers the three dimensionless parameters that govern the dynamics.
\begin{itemize}
    \item The parameter $\xi$ determines the Q factor of the oscillator in $\alpha(t)$ or $\beta(s)$.
    \item  The parameter $\frac{\Omega}{\Phi \omega_0}$ controls the capacity of the servomotor to perfectly follow the command $\psi_c(s)$. If $\frac{\Omega}{\Phi \omega_0}\gg 1$, we can eliminate adiabatically the dynamics of $\psi(s)$: in this limit we replace the Eq. for the wheel angle $\phi$  by $\phi(t)=\phi_c(t)$.
    This corresponds to the definition of a fast servomotor.
    \item The parameter $\Phi/\Delta$ measures the nonlinear response of the servomotor.
    \begin{itemize}
        \item if $\Phi/\Delta\gg 1$, the function $\tanh$, can be replaced by the function $\mathrm{sign}$ and because of the relaxational dynamics, the servomotor equation is reduced to %\textcolor{red}{J: To be as general as possible, wouldn't it be better to write $\mathrm{sign}  \left( \phi_c - \phi \right) $ ?. Unless we are considering $\phi_c = \Phi > |\phi| $ }
     %   \textcolor{blue}{M: comme ya une dyanmique de relxation, normalement il n'y a pas d'overshoot.}
        \begin{equation}
            \dot \phi (t)=\Omega \mathrm{sign}{\phi_c(t)} 
        \end{equation}
        \item if $\Phi/\Delta\ll 1$, the equation in $\psi$ can be linearized, and we can write the equation :
        \begin{equation}
            \dot \psi(s) =\frac{\Omega}{ \omega_0\Delta}\left(\psi_c(s)-\psi(s)\right),
        \end{equation}
        or in dimension variables:
        \begin{equation}
            \dot \phi(t)=\frac{\Omega}{\Delta} \left(\phi_c(t)-\phi(t)\right).
        \end{equation}
        Hence, the servomotor will perfectly follow the command if $\frac{\Omega}{ \omega_0\Delta}\gg  1$.
    \end{itemize}
\end{itemize}

In conclusion, the servomotor can be considered as fast in two regimes: first if $\frac{\Omega}{\Phi \omega_0}\gg 1$, second if $\frac{\Omega}{\Delta \omega_0}\gg 1$ and $\Phi/\Delta\ll 1$. 
% \textcolor{red}{Li: it seems that we do not need  $\Phi/\Delta$ to determine a fast servo ?}        \textcolor{blue}{M: pourquoi?}

\section*{Computations for fast servo motors}
In the limit of a fast servo motor, the variable $\phi$ and therefore $\alpha_c$ follow adiabatically $\phi_c$. 
Consequently, we set that the variable $\alpha_c \in [-\Lambda, \Lambda]$ is the control variable. 
%\sout{In this section, we detail the computations to derive the thrust for a biomimetic swimmer driven by a fast servomotor. }
In this limit, the  dynamics for the fin  angle $\alpha$ is described with:
\begin{eqnarray}
    \label{eq:dotalphaFast}
    \dot \alpha&=&\omega\\
    \label{eq:dotomegaFast}
    \dot \omega &=& -\xi\omega_0 \omega-\omega_0^2(\alpha-\alpha_c), \ |\alpha_c|\leq\Lambda
\end{eqnarray}
While the Hamiltonian $H$ from Eq. (\ref{eq:Hamiltonian}) reduces to 
\begin{eqnarray}
    H&=&-\omega^2+p_1 \omega
    +p_2\left(-\xi \omega_0 \omega-\omega_0^2(\alpha- \alpha_c)\right).
\end{eqnarray}
We remark that $H=\alpha_c(\omega_0^2p_2)+..$.
Hence, by setting $\alpha_c=-\Lambda$, if $p_2>0$ and $\alpha_c=\Lambda$, if $p_2<0$, we impose that the Hamiltonian is always minimum.
Consequently, if $p_2$ oscillates then the control $\alpha_c$ also oscillates between the two constant values extreme values $\pm \Lambda$.
The optimal command is therefore:
\begin{equation}
    \alpha_c=\Lambda \mathrm{sign}(p_2),
    \label{eq:control}
\end{equation}
which corresponds to the classical Bang-Bang controller.
We now study the dynamics of the $p_{1,2}$. The equations of the co-states are:
\begin{eqnarray}
    \dot p_1&=&-\partial_\alpha H=p_2 \omega_0^2\\
    \dot p_2&=&-\partial_\omega H= 2\omega-p_1+\xi \omega_0 p_2.
\end{eqnarray}
The equation for $p_2$ is written in the form:
\begin{equation}
    \label{eq:lambda2}
    \ddot p_2-\xi \omega_0 \dot p_2+\omega_0^2 p_2= 2\dot \omega, \ \ p_2(0)=0, \ \ p_2(T)=0
\end{equation}
which represents a linear oscillator forced by $\dot \omega$ and submitted to an injection of energy (the term proportional to $\dot p_2$).
As the equation for $\alpha$ of those of a harmonic oscillator if $\alpha_c$ is constant, the r.h.s. of eq. (\ref{eq:lambda2}) is harmonic. Consequently, the co-state $p_2$ should be a harmonic function whose amplitude grows in time because of the injection of energy. This demonstrates that $p_2$ oscillates, yielding oscillations in $\alpha_c$.

In the next part of this section we study the system (\ref{eq:dotalpha},\ref{eq:dotomega}).
We impose that $\alpha_c=\Lambda$ for $t \in [0, T/2]$ and $\alpha_c=-\Lambda$ for $t \in [T/2, T]$.

We solve the equation for the angle $\alpha$  in the first half period:
\begin{align}
    \alpha&=\Lambda+e^{-\frac{\xi \omega_0}{2}t}\left(a \cos\nu t+b\sin \nu t\right)\\
    \label{eq:nu}
    \nu&=\frac{1}{2} \omega_0 \sqrt{4-\xi^2},
\end{align}
where $a$ and $b$ are fixed by two boundary conditions.
Applying the condition of continuity and differentiability: 
\begin{align}
    \label{eq:bc1}
    \alpha(0)&=-\alpha(T/2)\\
    \label{eq:bc2}
    \dot \alpha(0)&=-\dot \alpha(T/2),
\end{align}
the unknowns $a$ and $b$ are determined:
\begin{align}
    a&=   \frac{\Lambda \xi \omega_0 \sin \left( \frac{T \nu}{2} \right) -2 \Lambda \nu \cos \left( \frac{T \nu}{2} \right)  -2 \Lambda \nu e^{T \xi \omega_0 / 4}  }{2 \nu 
   \left(\cosh \left(\frac{1}{4} \xi  T \omega _0\right)+\cos \left(\frac{T \nu }{2}\right)\right)}   \\
   b&=-\frac{\Lambda \xi \omega_0 \cos \left( \frac{T \nu}{2} \right)  + 2 \Lambda \nu \sin \left( \frac{T \nu}{2} \right) + 2 \Lambda \nu e^{T \xi \omega_0 / 4}   }{2
   \nu  \left(\cosh \left(\frac{1}{4} \xi  T \omega _0\right)+\cos \left(\frac{T \nu }{2}\right)\right)}.
\end{align}

We remark here that the above formulation is not sensitive to the sign of $\xi-2$, because the trigonometric functions become hyperbolic if their argument is imaginary.
It remains to compute the dynamics for $p_2$. In fact, the Eq. (\ref{eq:lambda2}) can be solved analytically, with the boundary conditions
\begin{align}
    p_2(0)&=0\\
    p_2(T/2)&=0\\
    \dot p_2(0)&=-\dot  p_2(T/2).
\end{align}
It appears that the last condition for the differentiability of $\lambda$ is automatically satisfied if $p_2$ is null at $t=0$ and $t=T/2$, because in Eq. (\ref{eq:lambda2}), the forcing $\omega$ is continuous. 

\begin{figure}[thb]
    \centering
    \includegraphics[width=0.95\textwidth]{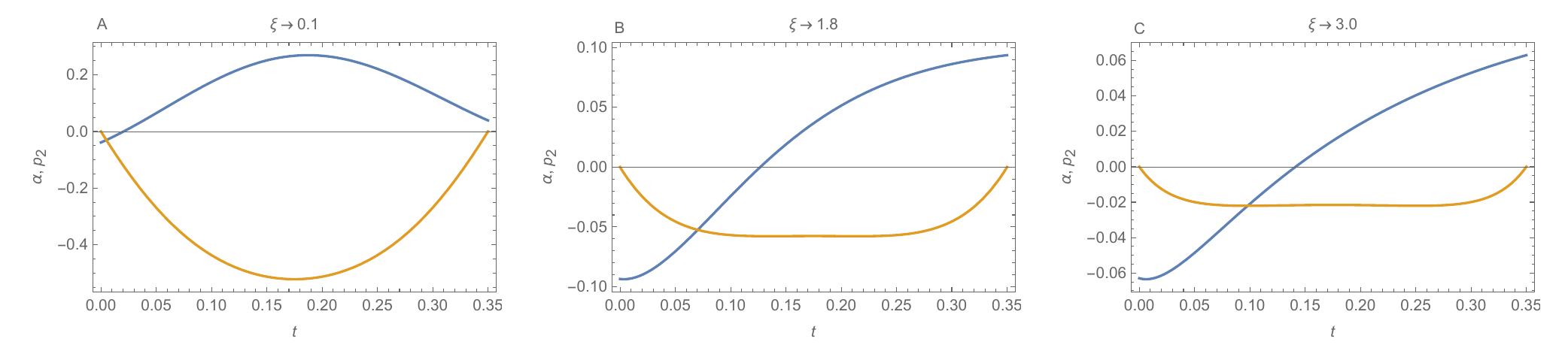}
    \caption{temporal evolution of $\alpha(t)$ (blue) and $p_2(t)$ (orange), obtained with $T=0.7$. $\omega_0=12.5\ \mathrm{s^{-1}}$, $\Lambda=0.1$}.
\label{fig:alpha_lambda}
\end{figure}

We show in Fig. \ref{fig:alpha_lambda}, various temporal evolution of $\alpha$ and $p_2$ in the interval $[0,T]$, for various values of the damping parameter $\xi$.

Consequently, the system provides oscillatory solutions independently of the chosen $T$.
%which is given as we search the optimal solution. 
We remark in the Fig. \ref{fig:alpha_lambda} that $p_2$ changes its sign as $\dot \alpha(t)$ does, such that a rule of thumb is to induce the change of $\dot\alpha_c$ as $\dot \alpha(t)$ ...is small enough, e.g. of order  $0.1\Phi \omega_0$.

\begin{figure}[htb]
\centering 
    \includegraphics[width=\textwidth]{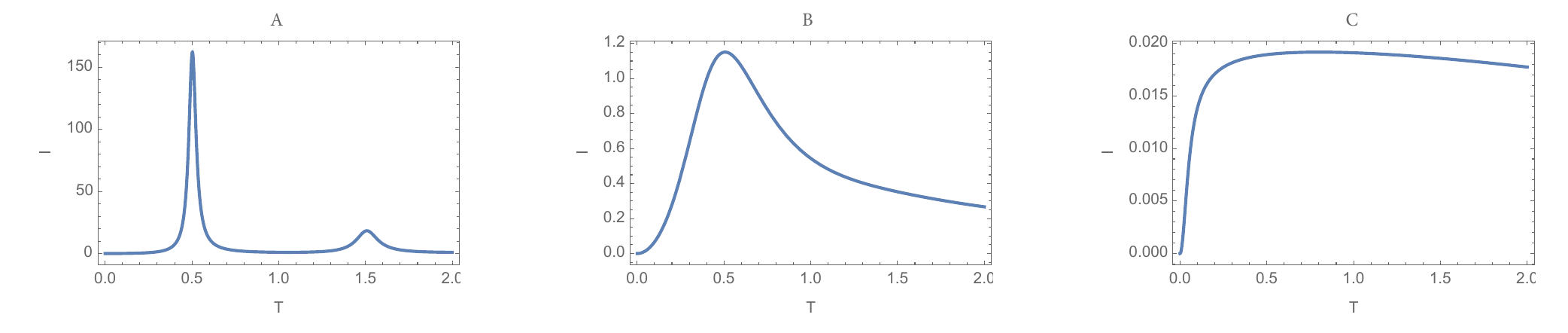}
\caption{Value of the dimensionless thrust as function of the period $T$, for various values of $\xi$ : a) $\xi=0.1$. b) $\xi=1.2$. c) $\xi=10$
}
\label{fig:Th}
\end{figure}

The average thrust writes:
\begin{align}
    I=&-\frac{1}{\Lambda^2 \omega_0^2}\frac{2}{T/2} \int_0^{T/2}\alpha \ddot \alpha dt\\ 
   =& \frac{4}{\xi \omega_0 T}\frac{2  \sinh \left(\frac{1}{4} \xi  T \omega _0\right)- \xi  \omega _0/\nu \sin \left(\frac{T \nu }{2}\right)}{
      \cos \left(\frac{T \nu }{2}\right)+  \cosh \left(\frac{1}{4} \xi  T
   \omega_0\right)}.
   \label{eq:Thrust}
\end{align}
The function $I$ presents a maximum near $T=\frac{2\pi}{\omega_0}$, as shown in Fig. \ref{fig:Th}.

This dimensionless thrust can be further optimized by computing the value $T^*$ that renders it maximal. We remark, at this point, that the parameter $T$ always appears multiplied by $\omega_0$ in Eq. \ref{eq:Thrust}, since $\nu$ is proportional to $\omega_0$ (as defined in Eq. \ref{eq:nu}):  $I$ only depends on $\xi$ and $\omega_0 T$.

\begin{figure}[htb]
\centering 
    \includegraphics[width=0.95\textwidth]{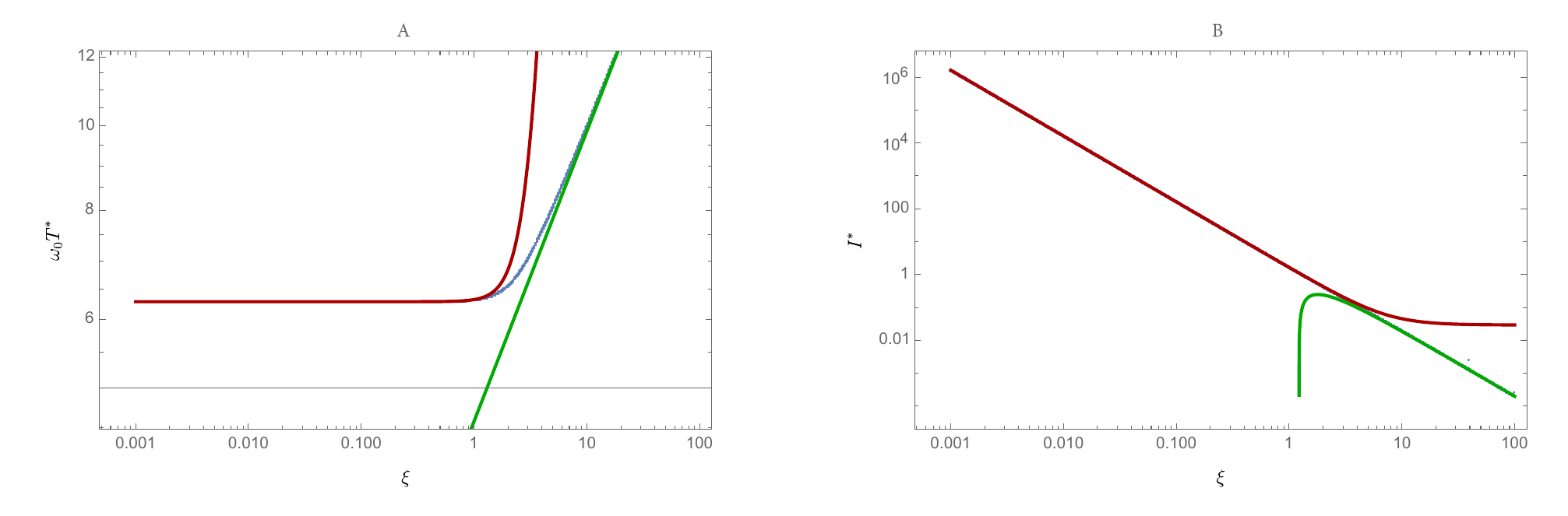}
\caption{a) Optimal dimensionless period $\omega_0 T^\star$ as function of $\xi$. 
b) Optimal dimensionless force $I^\star$ as function of $\xi$. The numerical solutions from Eq. (\ref{eq:Thrust}) are drawn in blue. The small damping limit is the red line and the large damping asymptotics are shown in green.
}
\label{fig:Ts_and_I_vs_Xi}
\end{figure}

In Fig. \ref{fig:Ts_and_I_vs_Xi}, we plot $\omega_0 T^*$  and $I^\star$ as a function of $\xi$, which corresponds to the optimal thrust for fast servomotors.

\subsection*{Limit of small damping: $\xi\rightarrow 0$}
In fact, assuming a small value for $\xi$, we show that the thrust is maximized at $T=T^\star$ defined by:
\begin{equation}
    T^\star=\frac{2\pi}{\omega_0}\left(1+\xi^4\frac{1}{384}\left(12-\pi^2\right)\right)+o(\xi^6),
\end{equation}
while the optimal thrust $I^\star$ writes:
\begin{equation}
    I^\star=\frac{16}{\pi ^2 \xi ^2}+\frac{\pi ^2-9}{3 \pi ^2}+o(\xi^2).
\end{equation}
In the Fig. \ref{fig:Ts_and_I_vs_Xi}, we show the influence of the damping factor $\xi$ on the optimal period and thrust.

In the limit of small damping, $\xi \ll 1$, the optimal thrust diverges. The reason is that in this limit the oscillator in $\alpha$ resonates with the forcing $\alpha_c$, since $\alpha_c$ change its sign on the period $2\pi/\omega_0$.
This is seen by taking the limit $\xi \rightarrow 0$ for the expressions $a$ and $b$:
\begin{align}
    a&=-\Lambda -\frac{\xi  \left(\Lambda  \left(T \omega _0-2 \sin \left(\frac{T \omega _0}{2}\right)\right)\right)}{4 \left(\cos \left(\frac{T \omega _0}{2}\right)+1\right)}+O\left(\xi ^2\right)\\
    b&=-\frac{\Lambda  \sin \left(\frac{T \omega _0}{2}\right)}{\cos \left(\frac{T \omega _0}{2}\right)+1}-\frac{\xi  \Lambda }{2}+O\left(\xi ^2\right),
\end{align}
The constant $a$ diverges as $T\rightarrow \frac{2\pi }{\omega_0}$, such that the swimming amplitude also diverges, and so the thrust.

In this limit we find that:
\begin{align}
    \alpha(0)&=-\frac{4 \Lambda }{\pi  \xi }-\frac{\xi  \left(\left(\pi ^2-9\right) \Lambda \right)}{12 \pi
   }+O\left(\xi ^3\right)\\
   \dot \alpha(0)&=-\frac{\Lambda  \omega_0}{\pi }+\frac{\xi ^2 \Lambda  \omega_0}{8 \pi
   }+O\left(\xi ^3\right)
\end{align}

\subsection*{Limit of large damping : $\xi \rightarrow \infty$}
In this limit, we get the following relation for the thrust:
$$
I=\frac{2}{\xi ^2}\left(1-\frac{4}{\xi  T \omega _0}+\frac{1}{\xi ^2}-\frac{T^2 \omega _0^2}{48 \xi ^2}\right)
$$
   The maximum of the thrust is obtained by differentiating the above formula with respect to $T$ and computing $T^\star$ that zeroes this derivative.
   For large $\xi$, we obtain the optimal period $T^\star$ and thrust $I^\star$:
   \begin{eqnarray}
    T^\star&=&\frac{1}{\omega_0}(96 \xi)^{1/3}+o(\xi^{1/3})\\
    I^\star&=&\frac{2}{\xi ^2}\left(1-\frac{\left(\frac{3}{2}\right)^{2/3}}{\xi ^{4/3}}\right)
   \end{eqnarray}

In this limit, the system is over-damped. A boundary layer appears near $t=0$, the size of this boundary layer is $1/(\xi \omega_0)$. This value is deduced by balancing the second derivative term with the damping term. Following the typical techniques for the asymptotic limit, we get
\begin{align}
    \ddot \alpha_i+\xi\omega_0 \dot \alpha_i=0\\
    \xi\omega_0 \dot \alpha_o+\omega_0^2(\alpha_o-\Lambda)=0\\
    \alpha(t)=\alpha_o(t)+\alpha_i(t)-\alpha_o(0),
\end{align}
where $\alpha_i(t)$ is the inner approximation of $\alpha(t)$ near $t=0$, i.e. the inner region (where the function is rapidly varying). $\alpha_o(t)$ the outer region, where the function is slowly varying.

We find that
\begin{align}
    \alpha(0)&=-\frac{\sqrt[3]{\frac{3}{2}}  }{\xi ^{2/3}}\Lambda\\
    \dot\alpha(0)&=-\frac{\Lambda \omega_0}{\xi}
\end{align}

\subsection*{Analysis of the swinging strategy}
\begin{figure}[htb]
\centering 
    \includegraphics[width=\textwidth]{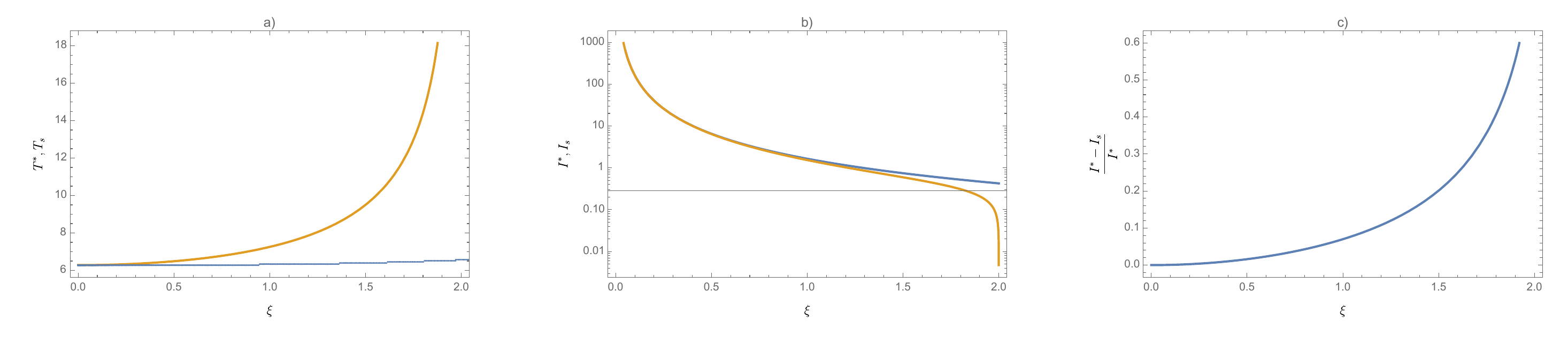}

    \caption{
        a) Comparison between $T^\star$ (blue) and $T_s$ (orange). 
        b) Comparison of the thrust $I$ obtained for $T=T^\star$ (blue) and $T=T_s$ (orange). 
        c) Relative error between the optimal thrust $I^*$ and the resulting thrust obtained by the swinging approach using $\dot \alpha(0)=0$.
        Here $\omega_0=12.5 \ \mathrm{s^{-1}}$, $\xi=1.2$ and $\Lambda=0.1$.
        }
    \label{fig:Ts}
\end{figure}

Here, we would like to measure the efficiency of the swinging strategy at least for $C=0$, where $C$ is defined in the main text. It consists in changing the sign of $\alpha_c$ as $\dot \alpha(t)$ zeroes. In the cruising regime, we have computed so far, this strategy predicts that $\dot \alpha(0)=\dot \alpha(T/2)$ should be zero.  
We therefore compute $\dot \alpha(0)$ to test the strategy:
$$
    \dot \alpha(0)=-\frac{2 \Lambda  \omega _0 \sin \left(\frac{1}{4} \sqrt{4-\xi ^2} T \omega _0\right)}{\sqrt{4-\xi ^2} \left(\cos
   \left(\frac{1}{4} \sqrt{4-\xi ^2} T \omega _0\right)+\cosh \left(\frac{1}{4} \xi  T \omega _0\right)\right)}.
$$
This expression predicts that $\dot \alpha(0)=0$ for $T=T_s$ and the thrust $I_s$:
\begin{eqnarray}
    T_s&=&\frac{4 \pi }{\sqrt{4-\xi ^2} \omega _0}\\
    I_s&=&\frac{2 \sqrt{4-\xi ^2}}{\pi  \xi }\coth \left(\frac{\pi  \xi }{2 \sqrt{4-\xi ^2}}\right)
\end{eqnarray}
In Figure (\ref{fig:Ts}), we compare the optimal period $T^\star$ with $T_s$, as well as the resulting thrusts $I$: It appears that the swinging strategy is very efficient in automatically choosing the optimal period for relatively low damping $\xi$.
Nevertheless, we remark that the resulting thrust obtained with $T=T_s$ is very close to the optimal one for $\xi<1.5$.
Hence, the swinging strategy appears to be very efficient in reaching the optimal thrust without knowing the values of the physical parameters $\omega_0$ and  $\xi$.

  \section*{Computations for slow servo motors}
  Here we detail the computations obtained for servomotor which can not follow the command and work at maximal angular velocity $\Omega$. In this limit, the Eq. (\ref{eq:dotphi}) becomes $\dot\phi=\pm \Omega$. If we assume a periodic solution, during the first half period, we deduce:
  \begin{equation}
    \phi(t)=-\frac{\Omega T}{4}+\Omega t.
  \end{equation}
  The integration constant has been determined by assuming $\phi(0)=-\phi(T/2)$. We then solve the equation for $\alpha$:
  \begin{equation}
    \alpha=\lambda \frac{- \xi  \Omega -  \Phi  \omega _0+  t \omega _0 \Omega }{\omega _0}+c_1 e^{\frac{1}{2} t \left(-\sqrt{\xi ^2-4} \omega _0-\xi  \omega _0\right)}+c_2
   e^{\frac{1}{2} t \left(\sqrt{\xi ^2-4} \omega _0-\xi  \omega _0\right)},
  \end{equation}
  where $c_1$ and $c_2$ are defined through the boundary conditions (\ref{eq:bc1},\ref{eq:bc2}). We otbain that the dimensionless thrust writes:
  \begin{align}  
  I&= \frac{2\Omega^2}{\Phi^2\omega_0^2}+2 \Omega ^3
  \frac{-\sqrt{\xi ^2-4} \left(\xi ^2-1\right)  \sinh \left(\frac{\xi  \Phi  \omega _0}{\Omega }\right)+\xi  \left(\xi ^2-3\right)  \sinh
   \left(\frac{\sqrt{\xi ^2-4} \Phi  \omega _0}{\Omega }\right)}{\xi  \sqrt{\xi ^2-4} \Phi ^3 \omega _0^3 \left(\cosh \left(\frac{\sqrt{\xi ^2-4} \Phi  \omega _0}{\Omega
   }\right)+\cosh \left(\frac{\xi  \Phi  \omega _0}{\Omega }\right)\right)}
   \label{eq:slowServo}
  \end{align}

    \subsection*{Limit of small damping}
    By taking the limit $\xi\rightarrow 0$ on (\ref{eq:slowServo}), we deduce:
    \begin{equation}
        I=\frac{4 \Phi ^2 \omega _0^2}{15 \Omega ^2},
    \end{equation}
such that $F_x=2K\lambda^2\Omega^2$, as claimed in the main text.
    \subsection*{Limit of large damping}
    By taking the limit $\xi\rightarrow \infty$ on (\ref{eq:slowServo}), we deduce:
    \begin{equation}
        I=\frac{2}{3\xi^2}
    \end{equation}

\section*{Why the swinging strategy yields a resonance?}
In the main text, we have hypothesized that an efficient way to drive the swimmer to its resonnance without the knowledge of the frequency modes of deformations. The idea is to impose that
\begin{equation}
    \alpha_c= \Lambda \mathrm{sign}(\dot \alpha).
    \label{eq:SwingingAlphac}
\end{equation}
With this command, the damped oscillator can be solved in each half plane of the phase portrait $(\alpha,\dot\alpha)$:
\begin{align}
    \ddot \alpha&+\xi\dot\alpha +\omega_0^2(\alpha-\Lambda)=0,\quad \dot\alpha>0\\
    \ddot \alpha&+\xi\dot\alpha +\omega_0^2(\alpha+\Lambda)=0,\quad \dot\alpha>0.
\end{align}
Assuming a periodic motion for which $\dot \alpha >0$ for $0<t<T/2$ and $\dot \alpha <0$ for $T/2<t<T$, the motion on the second half period is deduced from the the value of $\alpha $ in $0<t<T/2$:
\begin{align}
    \alpha&=\Lambda+e^{-\frac{\xi \omega_0}{2}t}\left(a \cos\nu t+b\sin \nu t\right)\\
    \label{eq:nu2}
    \nu&=\frac{1}{2} \omega_0 \sqrt{4-\xi^2},
\end{align}
where $a$ and $b$ are defined with the following boundary conditions:
\begin{align}
    \dot \alpha(0)&=\dot \alpha(T/2)=0\\
    \alpha(0)&=- \alpha(T/2).
\end{align}
The first conditions on $\dot \alpha$ yields
\begin{align}
    b&=\frac{a \xi  \omega _0}{2 \nu },\\
    T&=\frac{4 \pi }{\sqrt{4-\xi ^2} \omega _0}.
\end{align}
This last condition states that the forcing (\ref{eq:SwingingAlphac}) leads a periodic motion with a period equal to $2\pi/\nu$. This period corresponds to the critical one that produces the highest response in $\alpha$, i.e. a resonance in the amplitude.  We remark, that in the limit of small $\xi$, the swinging policy drives the swimmer to the optimal thrust, as the swinging period tends to the optimal $T^\star$.

\bibliography{referencesAllometry.bib}

  \begin{figure}[htb]
\centering 
    \includegraphics[width=0.45\textwidth]{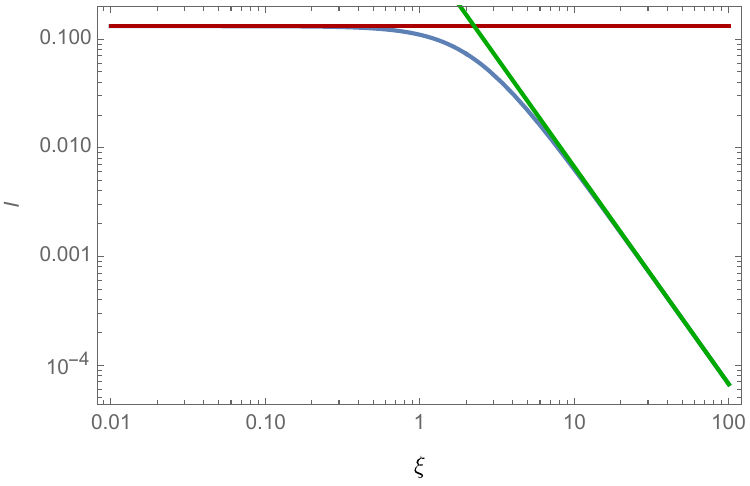}
    \caption{ Dimensionless thrust $I$ for a slow servomotor as function of the damping parameter $\xi$ (blue). We show the limits of small and large damping in red and green respectively.
        }
    \label{fig:Ts_slow}
\end{figure}

\clearpage